\documentclass{sig-alternate}

\usepackage{setspace}
\usepackage{multirow}
\usepackage{booktabs}
\usepackage{colortbl}
\usepackage{url}
\usepackage{paralist}
\usepackage{graphicx}
\usepackage{color}
\usepackage{minibox}
\usepackage{array}
\usepackage{booktabs}
\usepackage{verbatim}
\usepackage{amsmath}
\usepackage[table]{xcolor}
\usepackage{fancyvrb}
\usepackage{listings}
\usepackage{lipsum}
\usepackage{algorithm}
\usepackage{algpseudocode}
\usepackage{pifont}
\usepackage[export]{adjustbox}
\usepackage{soul}
\makeatletter

\def\therule{\makebox[\algorithmicindent][l]{\hspace*{.5em}\vrule height .75\baselineskip depth .25\baselineskip}}%

\newtoks\therules% Contains rules
\therules={}% Start with empty token list
\def\appendto#1#2{\expandafter#1\expandafter{\the#1#2}}% Append to token list
\def\gobblefirst#1{% Remove (first) from token list
	#1\expandafter\expandafter\expandafter{\expandafter\@gobble\the#1}}%
\def\LState{\State\unskip\the\therules}% New line-state
\def\pushindent{\appendto\therules\therule}%
\def\popindent{\gobblefirst\therules}%
\def\printindent{\unskip\the\therules}%
\def\printandpush{\printindent\pushindent}%
\def\popandprint{\popindent\printindent}%

\algdef{SE}[WHILE]{While}{EndWhile}[1]
{\printandpush\algorithmicwhile\ #1\ \algorithmicdo}
{\popandprint\algorithmicend\ \algorithmicwhile}%
\algdef{SE}[FOR]{For}{EndFor}[1]
{\printandpush\algorithmicfor\ #1\ \algorithmicdo}
{\popandprint\algorithmicend\ \algorithmicfor}%
\algdef{S}[FOR]{ForAll}[1]
{\printindent\algorithmicforall\ #1\ \algorithmicdo}%
\algdef{SE}[LOOP]{Loop}{EndLoop}
{\printandpush\algorithmicloop}
{\popandprint\algorithmicend\ \algorithmicloop}%
\algdef{SE}[REPEAT]{Repeat}{Until}
{\printandpush\algorithmicrepeat}[1]
{\popandprint\algorithmicuntil\ #1}%
\algdef{SE}[IF]{If}{EndIf}[1]
{\printandpush\algorithmicif\ #1\ \algorithmicthen}
{\popandprint\algorithmicend\ \algorithmicif}%
\algdef{C}[IF]{IF}{ElsIf}[1]
{\popandprint\pushindent\algorithmicelse\ \algorithmicif\ #1\ \algorithmicthen}%
\algdef{Ce}[ELSE]{IF}{Else}{EndIf}
{\popandprint\pushindent\algorithmicelse}%
\algdef{SE}[PROCEDURE]{Procedure}{EndProcedure}[2]
{\printandpush\algorithmicprocedure\ \textproc{#1}\ifthenelse{\equal{#2}{}}{}{(#2)}}%
{\popandprint\algorithmicend\ \algorithmicprocedure}%
\algdef{SE}[FUNCTION]{Function}{EndFunction}[2]
{\printandpush\algorithmicfunction\ \textproc{#1}\ifthenelse{\equal{#2}{}}{}{(#2)}}%
{\popandprint\algorithmicend\ \algorithmicfunction}%
\makeatother
\AtBeginDocument{}%

\copyrtyr={2015}
\copyrightetc{Copyright \the\copyrtyr\ by authors}

\begin{document}
%
% --- Author Metadata here ---
\conferenceinfo{ICSE}{'16 Austin, Texas USA (preprint)}
% --- End of Author Metadata ---

\title{SourcererCC: Scaling Code Clone Detection to Big Code}

\numberofauthors{1}
\author{
	\alignauthor
	Hitesh Sajnani\textsuperscript{*} Vaibhav Saini\textsuperscript{*} Jeffrey Svajlenko\textsuperscript{\textdagger} Chanchal K. Roy\textsuperscript{\textdagger}  Cristina V. Lopes\textsuperscript{*} \\
	\smallskip
	\affaddr{\textsuperscript{*}School of Information and Computer Science, UC Irvine, USA}\\
	\text{\{hsajani, vpsaini, lopes\}@uci.edu} \\
	\affaddr{\textsuperscript{\textdagger}Department of Computer Science, University of Saskatchewan, Canada}\\
	\text{\{jeff.svajlenko, chanchal.roy\}@usask.ca}
}

\maketitle

\begin{abstract}
Despite a decade of active research, there is a marked lack in clone
detectors that scale to very large repositories of source code, in
particular for detecting near-miss clones where significant editing
activities may take place in the cloned code. We present SourcererCC,
a token-based clone detector that targets three clone types,
and exploits an index to achieve scalability to large inter-project
repositories using a standard workstation. SourcererCC uses an
optimized inverted-index to quickly query the potential clones of a
given code block. Filtering heuristics based on token ordering are
used to significantly reduce the size of the index, the number of
code-block comparisons needed to detect the clones, as well as the
number of required token-comparisons needed to judge a potential
clone.
	
We evaluate the scalability, execution time, recall and precision of
SourcererCC, and compare it to four publicly available and
state-of-the-art tools. To measure recall, we use two recent
benchmarks, (1) a large benchmark of real clones, BigCloneBench,
and (2) a Mutation/Injection-based framework of thousands of
fine-grained artificial clones. We find SourcererCC has both high
recall and precision, and is able to scale to a large inter-project
repository (250MLOC) using a standard workstation.
\end{abstract}

\section{Introduction}
\label{sec:intro}

Clone detection locates exact or similar pieces of code, known as clones, within or between software systems.  Clones are created when developers reuse code by copy, paste and modify, although clones may be created by a number of other means~\cite{roy:queens:07}.  Developers need to detect and manage their clones in order to maintain software quality, detect and prevent new bugs, reduce development risks and costs, and so on~\cite{roy:queens:07, 6747168}. Clone management and clone research studies depend on quality tools.  According to Rattan et al.~\cite{Rattan20131165}, at least 70 diverse tools have been presented in the literature.

With the amount of source code increasing steadily, large-scale clone detection has become a necessity. Large-scale clone detection can be used for mining library candidates~\cite{ishihara}, detecting similar mobile applications~\cite{Chen:2014:android}, license violation detection~\cite{Koschke:CSMR:12,german:2009rb}, reverse engineering product lines~\cite{hemel:2012wcre,german:2009rb}, finding the provenance of a component~\cite{julius:2011:msr}, and code search~\cite{keivanloo:2011wcre,Kawaguchi:2009wcre}. Large-scale clone detection allows researchers to study cloning in large software ecosystems (e.g., Debian), or study cloning in open-source development communities (e.g., GitHub). Developers often clone modules or fork projects to meet the needs of different clients, and need the help of large-scale clone detectors to merge these cloned systems towards a product-line style of development. These applications require tools that scale to hundreds of millions of lines of code.  However, very few tools can scale to the demands of clone detection in very large code bases~\cite{jeff_scalability2, 6747168}.

A number of tools have been proposed to achieve a few specific applications of large-scale clone detection~\cite{Koschke:CSMR:12,keivanloo:2011wcre,Chen:2014:android}.  These tools make some assumptions regarding the requirements of their target domain that help with scalability.  These domain-specific tools are not described as general large-scale clone detectors, and may face significant scalability challenges for general clone detection.  General purpose clone detection is required for clone studies in large inter-project repositories and to help developers manage and merge their related software forks, as well as for use in the domain-specific activities.  Scalable general purpose clone detection has been achieved by using deterministic~\cite{livieri:2007icse} or non-deterministic~\cite{jeff_scalability2} input partitioning and distributed execution of an existing non-scalable detector, using large distributed code indexes~\cite{hummel:icsm:2010}, or by comparing hashes after Type-1/2 normalization~\cite{ishihara}.  These existing techniques have a number of limitations.  The novel scalable algorithms~\cite{hummel:icsm:2010,ishihara} do not support Type-3 near-miss clones, where minor to significant editing activities might have taken place in the copy/pasted fragments, and therefore miss a large portion of the clones, since there are more Type-3 in the repositories than other types~\cite{6747168,Roy:2010:NFC:1779593.1779596,bcb_icsme14}.  Type-3 clones can be the most needed in large-scale clone detection applications~\cite{6747168,keivanloo:2011wcre,Chen:2014:android}. While input partitioning can scale existing non-scalable Type-3 detectors, this significantly increases the cumulative runtime, and requires distribution over a large cluster of machines to achieve scalability in absolute runtime~\cite{livieri:2007icse,jeff_scalability2}. Distributable tools~\cite{livieri:2007icse} can be costly and difficult to setup. 

We set out to develop a clone detection technique and tool that would satisfy the following requirements: (1) accurate detection of near-miss clones, where minor to significant editing changes occur in the copy/pasted fragments; (2) programming language agnostic; (3) simple, non-distributed operation; and (4) scalability to hundreds of millions of lines of code.  To that effect, we introduce SourcererCC, a token-based accurate near-miss clone detector that exploits an optimized index to scale to hundreds of millions of lines of code (MLOC) on a single machine.  SourcererCC compares code blocks using a simple and fast bag-of-tokens\footnote{ Similar to the popular bag-of-words model~\cite{bag-of-words} in Information Retrieval} strategy which is resilient to Type-3 changes.  Clone candidates of a code block are queried from a partial inverted index.  A filtering heuristic is used to reduce the size of the index, which drastically reduces the number of required code block comparisons to detect the clones.  It also exploits the ordering of tokens to measure a live upper bound on the similarity of code blocks in order to reject or accept a clone candidate with fewer token comparisons. We found this technique has strong recall and precision for the first three clone types. SourcererCC is able to accurately detect exact and near-miss clones in 250MLOC on a single machine in only 4.5 days. We make two different versions of the tool available: (i) SourcererCC-B, a batch version of the tool that is more suitable for empirical analysis of the presence of clones in a system or a repository; and (ii) SourcererCC-I, an interactive version of the tool integrated with Eclipse IDE to help developers instantly find clones during software development and maintenance.

We evaluate the scalability, execution time and detection quality of SourcererCC. We execute it for inputs of various domains and sizes, including the Big Data inter-project software repository IJaDataset-2.0~\cite{ijadataset} (25,000 projects, 250MLOC, 3 million files), and observed good execution time and no scalability issues even 
on a standard machine with 3.5GHz quad-core i7 CPU and 12GB of memory. We measure its clone recall using two proven~\cite{jefficsme, bcb_icsme15} clone benchmarks.  We use BigCloneBench~\cite{bcb_icsme14}, a Big Data benchmark of real clones that spans the four primary clone types and the full spectrum of syntactical similarity.  We also use The Mutation and Injection Framework~\cite{mf_iwsc13, mf_icstw09}, a synthetic benchmark that can precisely measure recall at a fine granularity.  We measure precision by manually validating a sample of its output.  We compare these results against public available popular and state-of-the-art tools, including CCFinderX~\cite{ccfinder}, Deckard~\cite{deckard}, iClones~\cite{iclones} and NiCad~\cite{nicad}.  We find that SourcererCC is the only tool to scale to Big Data inputs without scalability issues on standard workstation. SourcererCC has strong precision and recall, and is competitive with the other tools.

\noindent \textbf{Outline:} The rest of the paper is organized as follows. Section~\ref{sec:definitions} describes important concepts and definitions. Section~\ref{sec:sourcererCC} presents SourcererCC's clone detection process in detail. Section~\ref{sec:evaluation} describes various experiments conducted to evaluate the scalability, recall and precision of SourcererCC against state-of-the-art tools on various benchmarks, with threats to validity discussed in Section~\ref{sec:threats}. After drawing connections with the related work in Section~\ref{sec:rel-work}, Section~\ref{sec:conclusion} concludes with a summary of the findings.  

\section{Definitions}
\label{sec:definitions}
The paper uses following well-accepted definitions of code clones and clone types~\cite{bellon,roy:queens:07}.:\\
\indent\textbf{Code Fragment}: A continuous segment of source code, specified by the triple $(l,s,e)$, including the source file $l$, the line the fragment starts on, $s$, and the line it ends on, $e$. \\
\indent\textbf{Clone Pair}:  A pair of code fragments that are similar, specified by the triple $(f_{1},f_{2},\phi)$, including the similar code fragments $f_{1}$ and $f_{2}$, and their clone type $\phi$.\\
\indent\textbf{Clone Class}: A set of code fragments that are similar.  Specified by the tuple $({f_{1},f_{2},...,f_{n}},\phi)$.  Each pair of distinct fragments is a clone pair: $(f_{i},f_{j},\phi),~i,j\in{1..n},~i\neq j$.

\indent \textbf{Code Block}: A sequence of statements, local class and variable declaration statements within braces.

\indent\textbf{Type-1(T1)}: Identical code fragments, except for differences in white-space, layout and comments. \\
\indent\textbf{Type-2(T2)}: Identical code fragments, except for differences in identifier names and literal values, in addition to Type-1 clone differences. \\
\indent\textbf{Type-3(T3)}: Syntactically similar code fragments that differ at the statement level.  The fragments have statements added, modified and/or removed with respect to each other, in addition to Type-1 and Type-2 clone differences.  \\
\indent\textbf{Type-4(T4)}: Syntactically dissimilar code fragments that implement the same functionality

%\subsection{Problem Formulation}

\section{The Proposed Method: S\lowercase{ourcerer}CC}
\label{sec:sourcererCC}
\subsection{Problem Formulation}
We assume that a project $P$ is represented as a collection of code blocks $P: \{B_1,...,B_n\}.$ In turn, a code block $B$ is represented as a bag-of-tokens $B: \{T_1...,T_k\}.$ A token is considered as programming language  keywords, literals, and identifiers. A string literal is split on whitespace and operators are not included. Since a code block may have repeated tokens, eack token is represented as a ($token, frequency$) pair. Here, $frequency$ denotes the number of times $token$ appeared in a code block.

In order to quantitatively infer if two code blocks are clones, we use a similarity function which measures the degree of similarity between code blocks, and returns a non-negative value. The higher the value, the greater the similarity between the code blocks. As a result, code blocks with similarity value higher than the specified threshold are identified as clones.

Formally, given two projects $P_x$ and $P_y$, a similarity function $f$, and a threshold $\theta$, the aim is to find all the code block pairs (or groups) $P_x.B$ and $P_y.B$ $s.t$ $f(P_x.B, P_y.B) \ge \lceil\:\theta\cdot\max(|P_x.B|,|P_y.B|)\:\rceil$. Note that for intra-project similarity, $P_x$ and $P_y$ are the same. Similarly, all the clones in a project repository can be revealed by doing a self-join on the entire repository itself. 
 
While there are many choices of similarity function, we use $Overlap$\footnote[3]{The presented approach can be used with other Jaccard and Cosine similarity functions as well.} because it intuitively captures the notion of overlap among code blocks. For example, given two code blocks $B_x$ and $B_y$, the overlap similarity $O(B_x, B_y)$ is computed as the number of source tokens shared by $B_x$ and $B_y$. 
 \begin{equation}
 	O(B_x, B_y) = |B_x \cap B_y|
 \end{equation} 
In other words, if $\theta$ is specified as $0.8$, and $max(|B_x|,|B_y|)$ is $t$, then $B_x$ and $B_y$ should at least share $\lceil \theta \dot{|t|} \rceil$ tokens to be identified as a clone pair\\

In order to detect all clone pairs in a project or a repository, the above approach of computing similarity between two code blocks can simply be extended to iterate over all the code blocks and compute pairwise similarity for each code block pair. For a given code block, all the other code blocks compared are called candidate code blocks or candidates in short. 

\begin{figure}
	\centering
	\includegraphics [scale=0.48]{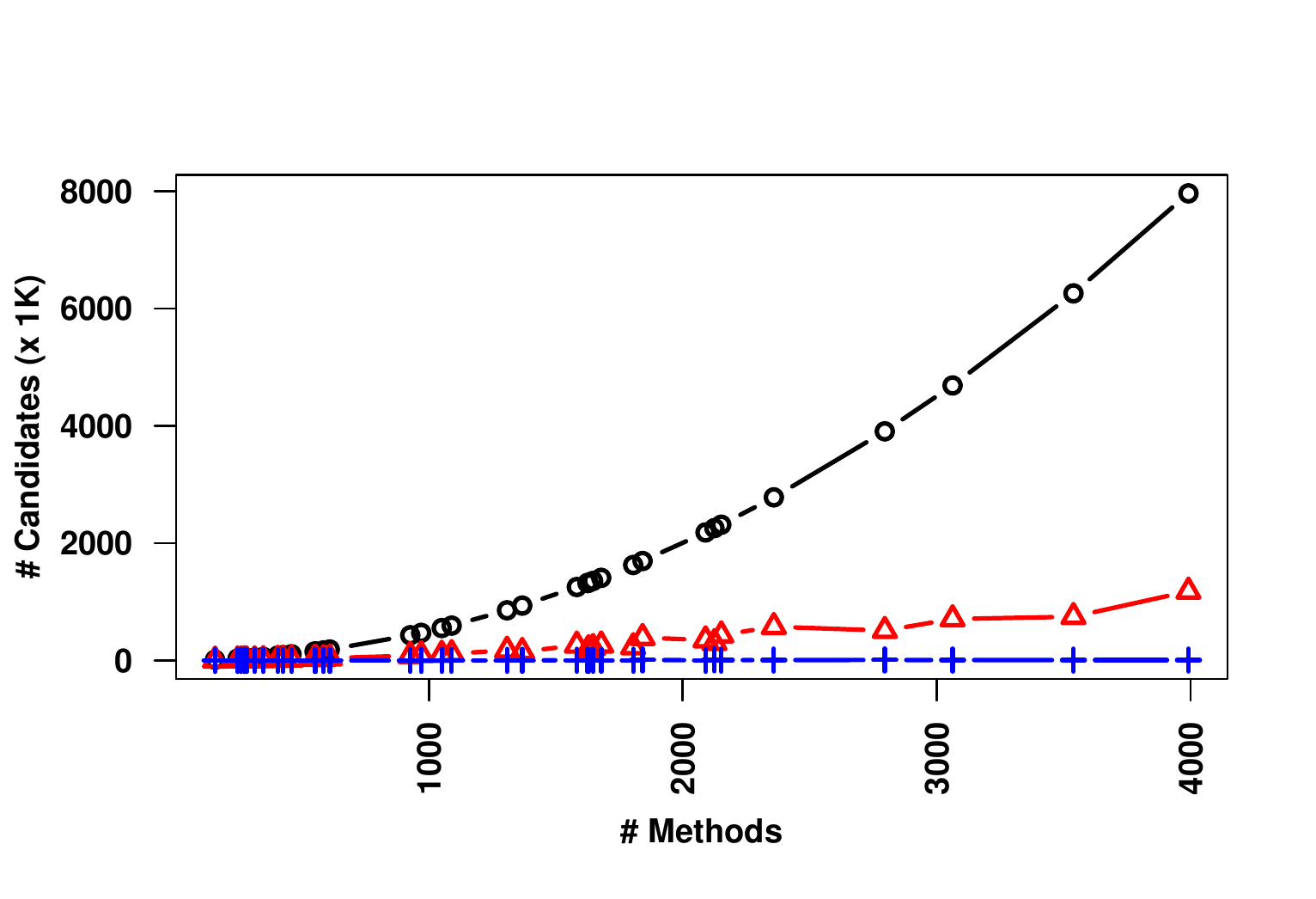}
	\caption {Growth in number of candidate comparisons with the increase in the number of code blocks}
	\label{fig:candidate_growth_pattern} 
\end{figure}

While the approach is very simple and intuitive, it is also subjected to a fundamental problem that prohibits scalability - $O(n^2)$ time complexity 
Figure~\ref{fig:candidate_growth_pattern} describes this by plotting the number of total code blocks (X-axis) vs. the number of candidate  comparisons (Y-axis) in $35$ Apache projects~\footnote{The list is available at \url{http://mondego.ics.uci.edu/projects/SourcererCC/}}. Note that the granularity of a code block is taken as a method. Points denoted by the $\color{black} \circ$ show that the number of candidate comparisons increase quadratically\footnote{The curve can also be represented using $y = x(x-1)/2$ quadratic function where $x$ is the number of methods in a project and $y$ is the number of candidate comparisons carried out to detect all clone pairs.} with the increase in number of methods. Later in Section~\ref{sec:sourcererCC} while describing SourcererCC, we will propose two filtering heurisitics that significantly reduce the number of candidate comparisons during clone detection.   

\subsection{Overview}
SourcererCC's general procedure is summarized in Figure 3.  It operates in two primary stages: (i) partial index creation; and (ii) clone detection.

In the index creation phase, it parses the code blocks from the source files, and tokenizes them with a simple scanner that is aware of token and block semantics of a given language~\footnote{Currently we have implemented this for Java, C and C\#, but can be easily extended to other languages}. From the code blocks it builds an inverted index mapping tokens to the blocks that contains them. Unlike previous approaches, it does not create an index of all tokens in the code blocks, instead it uses a filtering heuristic (Section~\ref{sec:subblock-filtering}) to construct a partial index of only a subset of the tokens in each block.

In the detection phase, SourcererCC iterates through all of the code blocks, retrieves their candidate clone blocks from the index. As per the filtering heuristic, only the tokens within the sub-block are used to query the index, which reduces the number of candidate blocks.  After candidates are retrieved, SourcererCC uses another filtering heuristic (Section~\ref{sec:token-pos-filtering}), which exploits ordering of the tokens in a code block to measure a live upper-bound and lower-bound of similarity scores between the query and candidate blocks. Candidates whose upper-bound falls below the similarity threshold are eliminated immediately without further processing. Similarly, candidates are accepted as soon as their lower-bound exceeds the similarity threshold. This is repeated until the clones of every code block are located. SourcererCC exploits symmetry to avoid detecting the same clone twice.

In the following sections, we provide a detailed description of the filtering heuristics and overall clone detection algorithm.

\begin{figure}
\centering
 \includegraphics[scale=0.3]{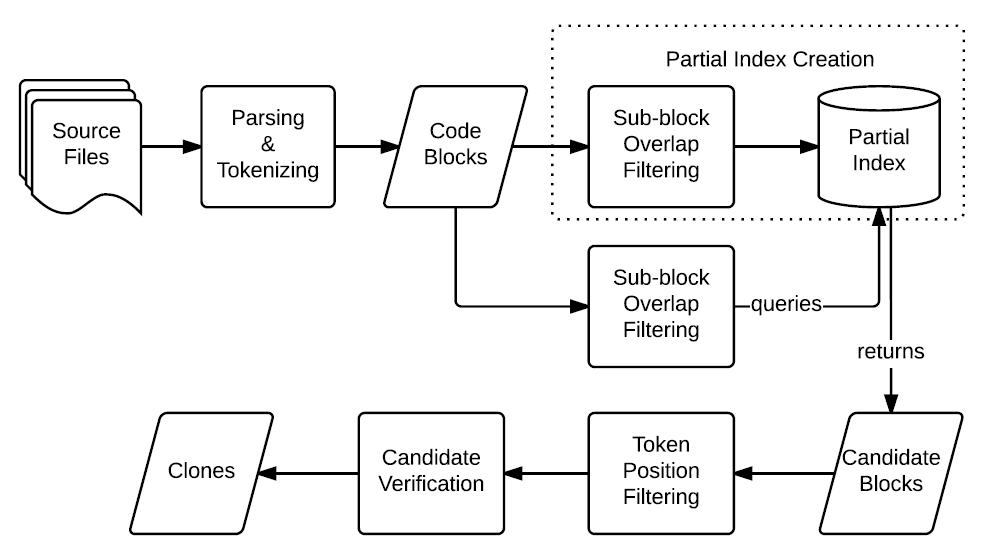} 
 \caption {SourcererCC's clone detection process}
\label{fig:clonealgo}
\centering
\end{figure}

\subsection{Filtering Heuristics to Reduce Candidate Comparisons}
This section describes filtering heurisitics that enable SourcererCC to effectively reduce the number of candidate code blocks comparison during clone detection.

\subsubsection{Sub-block Overlap Filtering}
\label{sec:subblock-filtering}
The filtering heuristics are inspired by the work of Sarawagi et
al.~\cite{sarawagi:2004} and Vernica et al.~\cite{vernica} on set
similarity. It follows an intuition that when two sets have a large
overlap, even their smaller subsets overlap. Since we represent code
blocks as bag-of-tokens (i.e. a multiset), we can extend this idea to
code blocks, i.e., when two code blocks have large overlap, even their
smaller sub-blocks should overlap as shown
in~\cite{sajnani:iwsc}. Formally, we can state it in the form of the
following property:

\noindent \textbf{Property 1}: {\emph{Given blocks $B_x$ and
    $B_y$ consisting of $t$ tokens each in some predefined order, if
    $|B_x\cap B_y| \ge i$, then the sub-blocks $SB_x$ and $SB_y$ of
    $B_x$ and $B_y$ respectively, consisting of first $t-i+1$ tokens,
    must match at least one token}.

To understand the implications of this property in clone detection,
let us consider two code blocks $B_x= \{\textbf{a, b}, c, d, e\}$ and
$B_y=\{\textbf{b, c}, d, e, f\}$ with $5$ tokens $(t=5)$ each. Let
$\theta$ be specified as $0.8$ meaning that the two blocks should
match at least $\lceil0.8*5\rceil = 4$ tokens to be considered clones
i.e, $(i=4)$.

According to Property 1, in order to find out if $B_x$ and $B_y$ are
clones, we can only check if their sub-blocks consisting of first
$t-i+1=2$ tokens match at least one token. In this case, they do, as
token $b$ is common in both the sub-blocks (marked in bold). However,
if they had not shared any token, then even without looking at the
remaining tokens of the blocks, we could have most certainly figured
that $B_x$ and $B_y$ will no t end up as a clone pair for the given
$\theta$. In other words, Property 1 suggests that instead of
comparing all the tokens of $B_x$ and $B_y$ against each other, we
could compare only their sub-blocks consisting of first $t-i+1$ tokens
to deduce if $B_1$ and $B_2$ will \emph{not} be clones.

In order to apply Property 1, the tokens in code blocks should follow a predefined global order. While there are many ways in which tokens in a block can be ordered e.g., alphabetical order, length of tokens, occurance frequency of token in a corpus, etc., a natural question is what order is most effective in this context. As it turns out, software vocabulary exhibits very similar characteristics to natural languages corpus and also follow Zipf's law~\cite{Hindle:2012,zipf}. That is, there are few very popular (frequent) tokens, and the frequency of tokens decreases very rapidly with rank. In other words, while most of the code blocks are likely to contain one or more of few very popular tokens (e.g., keywords, or common identifier names like $i$, $j$, $count$, etc.) not many will share rare tokens (e.g., identifiers that are domain or project specific). So if code blocks are ordered according to the popularity of tokens in the corpus, naturally, their sub-blocks will consist of these rare tokens. Such arrangement will ensure low probability of different sub-blocks sharing similar token. In other words, this ordering will eliminate more false positive candidates.\footnote{Candidates that eventually will not be identified as clones of a code block are known as false positive candidates for that block}.

To describe how effective this filtering is, points denoted by $\color{red} \triangle$ in Figure~\ref{fig:candidate_growth_pattern} show the number of candidate comparisons after applying the filtering. The difference with the earlier curve ($\color{black} \circ$) show the impact of filtering in eliminating candidate comparisons.   

The below section discusses when the use of Property 1 may still be ineffective and demonstrate how ordering of tokens in a code block can be further exploited to formalize yet another filtering heuristic that is extremely effective in eliminating even more candidate comparisons.

\subsubsection{Token Position Filtering}
\label{sec:token-pos-filtering}
In order to understand when Property 1 may be ineffective, consider code blocks $B_x$ and $B_y$ from the previous example, except $B_x$ now has one fewer token. Hence $B_x= \{\textbf{a, b}, c, d\}$ and $B_y=\{\textbf{b, c}, d, e, f\}$.

Assuming the same value of $\theta$, the blocks must still match tokens ($\lceil\:\theta\cdot\max(|B_x|,|B_y|)\:\rceil$ = $\lceil0.8*5\rceil = 4$) to be a clone pair. But since the two blocks have only $3$ tokens in common, they cannot be identified as a clone pair. However, note that their sub-blocks (shown in bold) consisting of first $t-i+1=2$ tokens still have a common token $b$. As a result, \emph{Property 1} is satisfied and $B_y$ will be identified as a candidate of $B_x$ although $B_x$ and $B_y$ eventually will not end up as a clone pair. In general, cases when the code blocks have fairly different sizes it is likely that they may result into false positives even after satisfying Property 1.

Interestingly, to overcome this limitation, the ordering of tokens in code blocks can be exploited. For example, if we closely examine the position of the matched token $b$ in $B_x$ and $B_y$, we can obtain an estimate of the maximum possible overlap between $B_x$ and $B_y$ as the sum of \emph{current matched tokens} and the \emph{minimum number of unseen tokens in}
$B_x$ $and$ $B_y$, i.e., $1 + min(2, 4) = 3$. Since this upper bound on
overlap is already smaller than the needed threshold of $4$ tokens, we can
safely reject $B_y$ as a candidate of $B_x$. Note that we can compute a safe upper bound (without violating the correctness) because the blocks follow a predefined order. The above heuristic can be formally stated as follows.

\noindent \textbf{Property 2}: \emph{Let blocks $B_x$ and $B_y$ be ordered and $\exists$ token $t$ at index $i$ in $B_x$, $s.t$ $B_x$ is divided in to two parts, where $B_x(first)$ = $B_x[1...(i-1)]$ and $B_x(second)$ = $B_x[i...|B_x|)]$ \\
\indent Now if $|B_x\cap B_y| \ge \lceil\:\theta\cdot\max(|B_x|,|B_y|)\:\rceil$, then $\forall$ $t \in B_x \cap B_y$, $|B_x(first)\cap B_y(first)|$ + $min(|B_x(second)|,|B_y(second)|) \ge \lceil\:\theta\cdot\max(|B_x|,|B_y|)\:\rceil$ }

To describe how effective this filtering is, points denoted by $\color{blue} +$ in Figure~\ref{fig:candidate_growth_pattern} show the number of candidate comparisons after applying this filtering. The reduction is so significant that empirically on this dataset, the function seems to be \emph{near-linear}. This is a massive reduction in comparison with the quadratic function shown earlier without any filtering.

Although both the filtering heuristics are independent of each other, they 
complement each other to effectively reduce more number of candidate comparisons together than alone.

The index data structure in conjunction with the above filtering heuristics 
form the key components of SourcererCC to achieve scalability. The next section describes the complete algorithm of SourcererCC.

\subsection{Clone Detection Algorithm} %for \\ Detecting Clones

The algorithm works in two stages: (i) Partial Index Creation; and (ii) Clone Detection. Each step has filtering heuristics directly embedded in it as described below.

\textbf{Partial Index Creation.} In traditional index based approaches, all the tokens are indexed. However, SourcererCC's index creation step exploits Property 1 and creates indexes for tokens only in sub-blocks. We call this \emph{Partial Index}. This not only saves space but also enables faster retrieval because of a smaller index.

Algorithm~\ref{alg:sourcererCCAlgoPIC} lists the steps to create a partial index. The first step is to iterate over each code block $b$ ($line\:3$), and sort it according to the global token frequency map $(GTP)$ ($line\:4$). This is done as a pre-requisite to the application of filtering based on Property 1. Next, the size of sub-block is computed using formula shown in Property 1 i.e., $(t-i+1)$. Later, tokens in the sub-block are indexed to create partial index. ($lines\:6-8$). 

\begin{algorithm}
	\scriptsize
		
		\textbf{INPUT:} $B$ is a list of code blocks \{$b_1$, $b_2$,...$b_n$\} in a project/repository, $GTP$ is the global token position map, and $\theta$ is the similarity threshold specified by the user \\
		\textbf{OUTPUT:} Partial Index($I$) of $B$\\
		
\begin{algorithmic}[1]
\Function{createPartialIndex}{B, $\theta$}
  	\LState $I$ = $\phi$
	\For{each code block $b$ in $B$}
			\LState $b$ = \Call{Sort}{$b$, $GTP$}		
			\LState $tokensToBeIndexed$ = $|b|-\lceil$ $\theta$ $\cdot{|b|}\rceil+1$ 
			\For{$i=1:tokensToBeIndexed$}
			    \LState $t$ = $b[i]$
				\LState $I_{t}$ = $I_{t}$ $\cup$ ${(t,i)}$
				%\tcc{Here $t$ is a term at $b_{1}[i]$
            \EndFor
	\EndFor
	\LState \Return $I$
\EndFunction 
\end{algorithmic}
\caption{SourcererCC's Algorithm - Partial Index Creation}
\label{alg:sourcererCCAlgoPIC}
\end{algorithm}

%\vspace{0.2cm}
\textbf{Clone Detection.} 
After partial index is created, the goal is to detect clones. Algorithm~\ref{alg:sourcererCCAlgoCD} describes the steps in detail. The \emph{detectClones()} function iterates over each query block $b$, and sorts them using the same $(GTP)$ that was created during index creation $(line\:4)$. Again, this is done as a prerequisite for both Property 1 \& 2 to be applicable. After that, it calculates the length of query sub-block by using the same formula described in Property 1 $(line\:5)$. Next it iterates over only as many tokens as the length of $b$'s sub-block and retrieves candidates by querying the partial index. Note that since partial index is created using only sub-blocks, the candidates retrieved in this phase implicitly satisfy Property 1. In other words, by creating partial index, the alogrithm not only reduces the index size, but also ensures that we only get filtered set of candidates that satisfy Property 1.

After the candidates are retrieved for a given query block, a trivial optimization to further eliminate candidates is done using size of the candidates. That is, if a candidate $c$ does not have enough tokens needed for it to be $b$'s clone pair, then there is no point in even comparing them. This is done using a conditional check $|c|$ $>$ $\lceil$ $\theta$  $\cdot$ $|b|$ $\rceil$ on $line\: 8$. This further filters out false positive candidates.

The remaining candidates that have satisfied the above elimination process are now subjected to the filtering based on Property 2. First, based on $\theta$, a threshold is computed that identifies the minimum number of tokens needed to be matched for $b$ and $c$ to be identified as a clone pair ($ct$ on $line\: 9$). Now, as the tokens in $b$ and $c$ are compared, a theoritical upper bound is dynamically computed based on the number of  remaining tokens in $b$ and $c$ ($line\:10$). This upper bound indicates the maximum number of tokens $b$ and $c$ could match assuming all of their tokens will match. If at any point in the iteration, the sum of upper bound (i.e, maximum number of tokens $b$ and $c$ could match) and the current similarity score (i.e, number of tokens $b$ and $c$ have matched) happens to be less than $ct$ (i.e, minimum number of tokens $b$ and $c$ need to match), $c$ is eliminated from $b$'s candidate map $candSimMap$ ($lines\:11$ $and$ $14$). In other words, it is violation of Property 2. On the other hand, if the sum is more than $ct$, the similarity between $b$ and $c$ gets updated with each token that is matched ($line\:12$). 
Once all the tokens in $b$'s sub-block are exhausted ($line\:19$), we have a map of candidates ($candSimMap$) along with their similarity score and the last seen token in each candidate. The reason for storing the last seen token will become clear as we explain futher. The next task is to verify if the candidates will eventually end up being $b$'s clones. This is done in a call to \emph{verifyCandidates()} function on $line\:18$.  

\textbf{Candidate Verification.}
The goal of \emph{verifyCandidates()} function is to iterate over  candidates $c$ of query $b$ that were not rejected in \emph{detectClones()}, compute their similarity score with $b$, and reject them if the score does not meet the computed threshold $ct$) or add them to the $cloneMap$ if it does.

In doing so, an important optimization is seen on $(line\:5)$. Note that tokens are not iterated from the start but from last token seen in $b$ and $c$ because earlier in \emph{detectClones()} few tokens of $b$ and $c$ were already iterated to check if they satisfy Property 1 \& 2 $(lines\: 6-8)$. Hence the function avoids iterating over those tokens again. It is for this reason, in \emph{detectClones()}, $candSimMap$ is designed to not only store candidates but also the last token that seen in each candidate, i.e., $(Candidate, TokensSeenInCandidate)$ pair. 

The rest of the function while iterating over the remaining tokens ensures that Property 2 holds at every iteration $(line\:6)$, and then increments the similarity score whenever there is a token match $(lines\:7-8)$. If at any iteration, Property 2 is violated, candidate is eliminated immediately without iterating over the remaining tokens $(line\:17)$. Thus saving much computation.

Another trivial but important optimization is done while iterating over code blocks. Since $b$ and $c$ are already sorted using a global token frequency (GTP), \emph{verifyCandidates()} efficienty iterates over $b$ and $c$ by incrementing only the index of a block that has a lower globally ranked token $(lines\: 10-14)$. Hence while iterating, except in the worst case when $b$ \& $c$ happen to be clone pairs, time complexity is reduced from $O(|b|*|c|)$ to $O(|b|+|c|)$.

\begin{algorithm}
	\scriptsize
		
		\textbf{INPUT:} $B$ is a list of code blocks \{$b_1$, $b_2$,...$b_n$\} in a project/repository, $I$ is the partial index created from $B$, and $\theta$ is the similarity threshold specified by the user 
		
		\textbf{OUTPUT:} All clone classes ($cloneMap$) \\
	
\begin{algorithmic}[1]
\Function{detectClones}{$B$, $I$, $\theta$}
	\For{each code block $b$ in $B$}
		\LState $candSimMap$ = $\phi$
		\LState $b$ = \Call{Sort}{$b$, $GTP$}
		\LState $querySubBlock$ = $|b|-\lceil$ $\theta$ $\cdot{|b|}\rceil+1$ 
		\For{$i=1:querySubBlock$}
			\LState $t$ = $b[i]$
			\For{each $(c,j)$ $\in$ $I_{t}$ such that $|c|$ $>$ $\lceil$ $\theta$  $\cdot$ $|b|$ $\rceil$}
				\LState $ct$ = $\lceil$ $max(|c|,|b|)$ $\cdot$ $\theta$ $\rceil$
				\LState $uBound$ = $1 + min(|b| - i, |c| - j)$
				\If{$candSimMap[c]$ $+$ $uBound$ $\ge$ $ct$}
					\LState $candSimMap[c]$ =  $candSimMap[c]+ (1, j)$
				\Else
					\LState $candSimMap[c]$ =  $(0, 0)$  \Comment{eliminate $c$}
				\EndIf
			\EndFor
		\EndFor
		\LState \Call{verifyCandidates}{$b,candSimMap,ct$}	
	\EndFor
	\LState \Return $cloneMap$ 
\EndFunction 
\end{algorithmic} 
\vspace{0.2cm}
\begin{algorithmic}[1]
\Function{verifyCandidates}{$b,candSimMap,ct$}
  	\For{each $c$ $\in$ $candSimMap$ such that $candSimMap[c] > 0$}
  	 	\LState $tokPos_c$  = Position of last token seen in $c$ 
  	  	\LState $tokPos_b$  = Position of last token seen in $b$ 
  	  	%\LState $sim$ = $simMap[c]$
  	 	
  	 	\While{$tokPos_b$ $<$ $|b|$ \&\& $tokPos_c$ $<$ $|c|$ }
  	 	\If {$min(|b|-tokPos_b$, $|c|-tokPos_c$) $\geq$ $ ct$}	
  	 			 \If{$b[tokPos_b]==c[tokPos_c]$}
  	 			 	\LState $candSimMap[c] = candSimMap[c]+1$
  	 			 \Else
  	 			 	\If{$GTP[b[tokPos_b]]<GTP[c[tokPos_c]]$}
  	 			 		\LState $tokPos_b++$
  	 			 	\Else
  	 			 		\LState $tokPos_c++$  
  	 			 	\EndIf
  	 			 \EndIf
  	 	\Else 
  	 		\LState break 
  	 	\EndIf 		 	
  		\EndWhile
  		\If {$candSimMap[c]$ $>$ $ct$}
  			\LState $cloneMap[b]$ = $cloneMap[b]$ $\cup$ $c$
  		\EndIf
  	\EndFor
 \EndFunction
\end{algorithmic}
\caption{SourcererCC's Algorithm - Clone Detection}
\label{alg:sourcererCCAlgoCD}
\end{algorithm} 

\subsection{Detection of Near-miss (Type-3) clones}
One of the distinguishing characteristics of SourcererCC compared to other token-based tools is its ability to detect Near-miss (Type-3) clones. The bag-of-tokens model plays an important role in this. Type-3 clones are created by adding, removing or modifying statements in a duplicated code fragment. Since the bag-of-tokens model is agnostic to relative token positions in the code block, it is resilient to such changes, and hence can detect near-miss clones as long as the code blocks (bags) share enough tokens to exceed a given overlap threshold.  

Many Type-3 clones have modifications of kind similar to swapping statement positions in code blocks, combining multiple condition expressions into one, changing operators in conditional statements, and use of one language construct over another (for vs while). While these changes may exhibit semantic difference, they preserve enough syntactic similarity at a token level to be detected as similar. Detecting such clones can be difficult for other token-based approaches as they use token sequences as a unit of match~\cite{ccfinder}. While a token-sequence approach could merge nearby cloned sequences into Type-3 clones~\cite{iclones}, they fail to detect the clones when the Type-3 gaps are too frequent or large.

\section{Evaluation}
\label{sec:evaluation}
In this section we evaluate the execution and detection performance of SourcererCC.  We begin by evaluating its execution time and scalability using subject inputs of varying sizes in terms of lines of code (LOC).  We then demonstrate SourcererCC's execution for a Big Data inter-project repository, one of the prime targets of scalable clone detection.  We measure its clone recall using two benchmarks: The Mutation and Injection Framework~\cite{mf_icstw09,mf_iwsc13} and BigCloneBench~\cite{bcb_icsme14, bcb_icsme15}.  We measure the precision of our tool by manually validating a statistically significant sample of its output for the BigCloneBench experiment.

We compare SourcererCC's execution and detection performance against four publicly available clone detection tools, including CCFinderX~\cite{ccfinder}, Deckard~\cite{deckard}, iClones~\cite{iclones} and NiCad~\cite{nicad}.  We include CCFinderX as it is a popular and successful tool, which has been used in many clone studies.  We include Deckard, iClones and NiCad as popular examples of modern clone detection tools that support Type-3 clone detection.  While we have benchmarked a number of tools in our previous work~\cite{jefficsme, bcb_icsme15}, we focus on those with the best scalability, recall, and/or most unique performance aspects for this study.  We focus primarily on near-miss clone detectors, as Type-1 and Type-2 clones are relatively easy to detect.  The configurations of these tools for the experiments are found in Table~\ref{tab:tool_configuration}.  These are targeted configurations for the benchmarks, are based on our extensive previous experiences~\cite{jefficsme, bcb_icsme15} with the tools, as well as previous discussions with their developers, where available.

\begin{table}[t]
	\centering
	\scriptsize
	\tabcolsep=0.05cm
	\caption{Clone Detection Tool Configurations}
	\label{tab:tool_configuration}
	\begin{tabular}{c>{\raggedright\arraybackslash}p{3.25cm}>{\raggedright\arraybackslash}p{3.25cm}}%>{\raggedright\arraybackslash}p{5cm}}
		\toprule
		Tool & \multicolumn{1}{c}{Scale/BigCloneBench} & \multicolumn{1}{c}{Mutation Framework} \\ % & \multicolumn{1}{c}{Bellon's Benchmark} \\
		\midrule
		SourcererCC & Min length 6 lines, min similarity 70\%, function granularity. & Min length 15 lines, min similarity 70\%, function granularity. \\ %& Min length 6 lines, min similarity 70\%, block granularity. \\
		%\midrule
		CCFinderX & Min length 50 tokens, min token types 12. & Min length 50 tokens, min token types 12. \\ % & Min length 25 tokens, min token types 6.\\
		Deckard  & Min length 50 tokens, 85\% similarity, 2 token stride. & Min length 100 tokens, 85\% similarity, 4 token stride. \\ % & Min length 30 tokens, 5 token stride, min 90\% similarity.\\
		iClones & Min length 50 tokens, min block 20 tokens. & Min length 100 tokens, min block 20 tokens. \\ % & Min length 30 tokens, min block 10 tokens.\\
		NiCad & Min length 6 lines, blind identifier normalization, identifier abstraction, min 70\% similarity. & Min length 15 lines, blind identifier normalization, identifier abstraction, min 70\% similarity. \\ % & Min length 4 lines, blind identifier normalization, identifier abstraction, min 70\% similarity.\\
		%ConQat~\cite{conqat}    & Min length 6 lines, max errors 5, gap ratio 30\%. & Min length 15 lines, max errors 3, gap ratio 30\%. \\
		%CPD~\cite{cpd} & Min length 50 tokens, ignore annotations/identifiers/literals, skip parser errors. & Min length 100 tokens, ignore annotations/identifiers/literals, skip parser errors.\\
		%CtCompare~\cite{ctcompare} & Min length 50 tokens, max 6 isomorphic relations. & Min length 100 tokens, max 3 isomorphic relations.\\
		%Duplo~\cite{duplo}     & Min length 6 lines.  Min 1 character per line. & Min length 15 lines.  Min 1 character per line. \\
		%SimCad~\cite{simcad}    & Greedy transformation, unicode support, min 6 lines. & Greedy transformation, unicode support, min 15 lines. \\
		%Simian~\cite{simian}   & Min length 6 lines, ignore identifiers and literals. & Min length 15 lines, ignore identifiers and literals. \\
		\bottomrule
	\end{tabular}
	\vspace{-0.2in}
\end{table}

Our primary goal with SourcererCC is to provide a clone detection tool that scales efficiently for large inter-project repositories with near-miss Type-3 detection capability.  Most existing state-of-the-art tools have difficulty with such large inputs, and fail due to scalability limits~\cite{jeff_scalability2,jeff_scalability1}.  Common  limits include untenable execution time, insufficient system memory, limitations in internal data-structures, or crashing or reporting an error due to their design not anticipating such a large input~\cite{jeff_scalability2,jeff_scalability1}.  We consider our tool successful if it can scale to a large inter-project repository without encountering these scalability constraints while maintaining a clone recall and detection precision comparable to the state-of-the-art.  As our target we use IJaDataset 2.0~\cite{ijadataset}, a large inter-project Java repository containing 25,000 open-source projects (3 million source files, 250MLOC) mined from SourceForge and Google Code.

\subsection{Execution Time and Scalability}
\label{sec:scale_eval}
In this section we evaluate the execution time and scalability of SourcererCC and compare it to the competing tools.  Execution time primarily scales with the size of the input in terms of the number of lines of code (LOC) needed to be processed and searched by the tool.  So this is the ideal input property to vary while evaluating execution performance and scalability.  However, it is difficult to find subject systems that are large enough and conveniently dispersed in size.  Additionally, a tool's execution time and memory requirements may also be dependent on the clone density, or other properties of the subject systems.  It is difficult to control for these factors while measuring execution performance and scalability in terms of input size.

Our solution was to build inputs of varying convenient sizes by randomly selecting files from IJaDataset.  This should ensure each input has similar clone density, and other properties that may affect execution time, except for the varying size in LOC.  Each input has the properties of an inter-project repository, which is a target of large-scale clone detection.  We created one input per order of magnitude from 1KLOC to 100MLOC.  We built the inputs such that each larger input contains the files of the smaller inputs.  This ensures that each larger subset is a progression in terms of execution requirements.  Lines of code was measured using cloc~\cite{cloc}, and includes only lines containing code, not comment or blank lines.

The execution time of the tools for these inputs is found in Table~\ref{tab:tool_configuration}.  The tools were executed for these inputs using the configurations listed under ``Scale'' in Table~\ref{tab:tool_scale}.  We limited the tools to 10GB of memory, a reasonable limit for a standard workstation.  The tools were executed on a machine with a 3.5GHz quad-core i7 CPU, 12GB of memory, a solid-state drive, and running Ubuntu 15.04.  CCFinderX was executed on an equivalent machine running Windows 7.  We use the same configurations for evaluating recall with BigCloneBench such that recall, execution performance and scalability can be directly compared.

\begin{table*}[t]
	\centering
	\scriptsize
	\tabcolsep=0.5cm
	\caption{Execution Time (or Failure Condition) for Varying Input Size}
	\label{tab:tool_scale}
	\begin{tabular}{cccccc}
		\toprule
		LOC   & SourcererCC & CCFinderX 			& Deckard 										& iClones 											& NiCad 									\\
		\midrule
		1K    & 3s              & 3s        		& 2s       										& 1s        										& 1s      									\\
		10K   & 6s              & 4s        		& 9s       										& 1s        										& 4s      									\\
		100K  & 15s             & 21s       		& 1m 34s        								& 2s        										& 21s      									\\
		1M    & 1m 30s          & 2m 18s   	 		& 1hr 12m 3s        							& \cellcolor[gray]{0.8}\color{black}{MEMORY}        & 4m 1s      								\\
		10M   & 32m 11s         & 28m 51s           & \cellcolor[gray]{0.8}\color{black}{MEMORY}    & ---        										& 11hr 42m 47s      						\\
		100M  & 1d 12h 54m s5s  & 3d 5hr 49m 11s	& ---        									& ---        										& \cellcolor[gray]{0.8}\color{black}{INTERNAL LIMIT} \\
		\bottomrule
	\end{tabular}
	\vspace{-0.2in}
\end{table*}

\textbf{Scalability.} SourcererCC is able to scale even to the largest input, with reasonable execution time given the input sizes.  CCFinderX is the only competing tool to scale to 100MLOC, however it only detects Type-1 and Type-2 clones.  The competing Type-3 tools encounter scalability limits before the 100MLOC input.  Deckard and iClones run out of memory at the 100MLOC and 1MLOC inputs, respectively.  NiCad is able to scale to the 10MLOC input, but refuses to execute clone detection on the 100MLOC input.  In our previous experience~\cite{jeff_scalability1}, NiCad refuses to run on inputs that exceeds its internal data-structure limits, which prevent executions that will take too long to complete.  From our experiment, it is clear that the state-of-the-art Type-3 tools do not scale to large inputs, whereas SourcererCC can.

\textbf{Execution Time.}  For the 1KLOC to 100KLOC inputs, SourcererCC has comparable execution time to the competing tools.  iClones is the fastest, but it hits scalability issues (memory) as soon as the 1MLOC input.  SourcererCC has comparable execution time to CCFinderX and NiCad for the 1MLOC input, but is much faster than Deckard.  SourcererCC has comparable execution time to CCFinderX for the 10MLOC input size, but is much faster than NiCad.  For the largest input size, SourcererCC is twice as fast as CCFinderX, although their execution times fall within the same order of magnitude.  Before the 100MLOC input, SourcererCC and CCFinderX have comparable execution times.

SourcererCC is able to scale to inputs of at least 100MLOC.  Its execution time is comparable or better than the competing tools.   Of the examined tools, it is the only state-of-the-art Type-3 clone detector able to scale to 100MLOC.  While CCFinderX can scale to 100MLOC for only detecting Type-1 and Type-2 clones, SourcererCC completes in half the execution time while also detecting Type-3 clones.

\subsection{Experiment with Big IJaDataset}
Since SourcererCC scaled to 100MLOC without issue, we also executed for the entire IJaDataset (250MLOC).  This represents the real use case of clone detection in a Big Data inter-project software repository.  We execute the tool on a standard workstation with a quad-core i7 CPU, 12GB of memory and solid state drive.  We restricted the tool to 10GB of memory and 100GB of SSD disk space.  We executed SourcererCC using the ``Scale'' configuration in Table~\ref{tab:tool_configuration}, with the exception of increasing the minimum clone size to ten lines.  Six lines is common in benchmarking~\cite{bellon}.  However, a six line minimum may cause an excessive number of clones to be detected in IJaDataset, and processing these clones for a research task can become another difficult scalability challenge~\cite{jeff_scalability2}.  Additionally, larger clones may be more interesting since they capture a larger piece of logic, while smaller clones may be more spurious.

SourcererCC successfully completed its execution for IJaDataset in 4 days and 12 hours, detecting a total of 146 million clone pairs.  The majority of this time was clone detection.  Extracting and tokenizing the functions required 3.5 hours, while computing the global token freqeuncy map and tokenizing the blocks required only 20 minutes.  SourcererCC required 8GB of disk space for its pre-processing, index (1.2GB) and output.  Of the 4.7 million functions in IJaDataset greater than 10 lines in length, 2.4 million (51\%) appeared in at least one clone pair detected by SourcererCC.  We have demonstrated that SourcererCC scales to large inter-project repositories on a single machine with good execution time.  We have also shown that building an index is an inexpensive way to scale clone detection and reduce overall execution time.

Since CCFinderX scales to the 100MLOC sample, we also executed it for IJaDataset.  We used the same settings as the scalability experiment.  We did not increase CCFinderX's minimum clone size from 50 tokens, which is roughly 10 lines (assuming 5 tokens per line).  This was not an issue with benchmarking as we used a 50 token minimum for reference clones from BigCloneBench.  CCFinderX executed for 2 days before crashing due to insufficient disk space.  Its pre-processed source files (25GB) and temporarily disk space usage (65GB) exceeded the 100GB reserved space.  Based on the findings of a previous study, where CCFinder was distributed over a cluster of computers~\cite{livieri:2007icse}, we can estimate it would require 10s of days to complete detection on 250MLOC, given sufficiently large disk-space.  So we can confidently say that SourcererCC is able to complete sooner, while also detecting Type-3 clones.

\subsection{Recall}
In this section we measure the recall of SourcererCC and the competing tools.  Recall has been very difficult for tool developers to measure as it requires knowledge of the clones that exist in a software system~\cite{roy:queens:07, 6747168}.  Manually inspecting a system for clones is non-trivial.  Even a small system like Cook, when considering only function clones, has almost a million function pairs to inspect~\cite{Walenstein:2003:PCT:950792.951349}.  Bellon et al.~\cite{bellon} created a benchmark by validating clones reported by the clone detectors themselves.  This has been shown to be unreliable for modern clone detectors~\cite{jefficsme}.  Updating this benchmark to evaluate your tool would require extensive manual clone validation with a number of modern tools.  As such, many clone detection tool papers simply do not report recall. 

In response we created The Mutation and Injection Framework~\cite{mf_icstw09,mf_iwsc13}, a synthetic benchmark that evaluates a tool's recall for thousands of fine-grained artificial clones in a mutation-analysis procedure.  The framework is fully automatic, and requires no validation efforts by the tool developer.  However, we recognized that a modern benchmark of real clones is also required.  So we developed an efficient clone validation strategy based on code functionality and built BigCloneBench~\cite{bcb_icsme14}, a Big Data clone benchmark containing 8 million validated clones within and between 25,000 open-source projects.  It measures recall for an extensive variety of real clones produced by real developers.  The benchmark was designed to support the emerging large-scale clone detection tools, which previously lacked a benchmark.  This combination of real-world and synthetic benchmarking provides a comprehensive view of SourcererCC's clone recall.

\subsubsection{Recall Measured by The Mutation Framework}

The Mutation Framework evaluates recall using a standard mutation-analysis procedure.  It starts with a randomly selected real code fragment (a function or a code block).  It mutates this code fragment using one of fifteen clone-producing mutation operators.  Each mutation operator performs a single code edit corresponding to one of the first three clone types, and are based on an empirically validated taxonomy of the types of edits developers make on copy and pasted code.  This artificial clone is randomly injected into a copy of a subject system.  The clone detector is executed for this system, and its recall measured for the injected clone.  The framework requires the tool to not only sufficiently report the injected clone, but appropriately handle the clone-type specific change introduced by the mutation operator.  As per mutation-analysis, this is repeated thousands of times.  Further details, including a list of the mutation operators, is available in our earlier studies~\cite{mf_icstw09, taxonomy, mf_iwsc13}.

\textbf{Procedure.} We executed the framework for three programming languages: Java, C and C\#, using the following configuration.  For each language, we set the framework to generate clones using 250 randomly selected functions, 10 randomly selected injection locations, and the 15 mutation operators, for a total of 37,500 unique clones per language (112,500 total).  For Java we used JDK6 and Apache Commons as our source repository and IPScanner as our subject system.  For C we used the Linux Kernel as our repository and Monit as our subject system.  For C\# we use Mono and MonoDevelop as our repository, and MonoOSC as our subject system.  We constrained the synthesized clones to the following properties: (1) 15-200 lines in length, (2) 100-2000 tokens in length, and (3) a mutation containment of 15\%.  We have found this configuration provides accurate recall measurement~\cite{jefficsme, bcb_icsme15}.  The tools were executed and evaluated automatically by the framework using the configurations listed in Table~\ref{tab:tool_configuration}.  To successfully detect a reference (injected) clone, a tool must report a candidate clone that subsumes 70\% of the reference clone by line, and appropriately handles the clone-type specific edit introduced by the mutation operator~\cite{mf_iwsc13}.

\textbf{Results.} Recall measured by the Mutation Framework for SourcererCC and the competing tools is summarized in Table~\ref{tab:mutationframework}.  Due to space considerations, we do not show recall per mutation operator.  Instead we summarize recall per clone type.  SourcererCC has perfect recall for first three clone types, including the most difficult Type-3 clones, for Java, C and C\#.  This tells us that it's clone detection algorithm is capable of handling all the types of edits developers make on copy and pasted code for these languages, as outlined in the editing taxonomy for cloning~\cite{taxonomy}.

\begin{table}[t]
	\centering
	\scriptsize
	\tabcolsep=0.125cm
	\centering
	\caption{Mutation Framework Recall Results}
	\label{tab:mutationframework}
	\begin{tabular}{cccccccccccc}
		\toprule
		\multirow{2}{*}{Tool} & \multicolumn{3}{c}{Java} & & \multicolumn{3}{c}{C} & & \multicolumn{3}{c}{C\#} \\
		\cmidrule{2-4}\cmidrule{6-8}\cmidrule{10-12}
		& T1 & T2 & T3 & & T1 & T2 & T3 & & T1 & T2 & T3 \\
		\midrule
		SourcererCC                 & 100 & 100 & 100          & & 100 & 100 & 100       & & 100 & 100 & 100         \\
		\midrule                   
		CCFinderX                   &  99 &  70 &   0          & & 100 &  77 &   0       & &  100 &  78 &   0         \\
		Deckard                     &  39 &  39 &  37          & &  73 &  72 &  69       & &   - &   - &   -         \\
		iClones                     & 100 &  92 &  96          & &  99 &  96 &  99       & &   - &   - &   -         \\
		NiCad                       & 100 & 100 & 100          & &  99 &  99 &  99       & &  98 &  98 &  98         \\
		\bottomrule
	\end{tabular}
	\vspace{-0.2in}
\end{table}

SourcererCC exceeds the competing tools with the Mutation Framework.  The runner up is NiCad, which has perfect recall for Java, and near-perfect recall for C and C\#.  iClones is also competitive with SourcererCC, although iClones has some troubles with a small number of Type-2 and Type-3 clones.  SourcererCC performs much better for Type-2 and Type-3 clones than CCFinderX.  Of course, as a Type-2 tool, CCFinderX does not support Type-3 detection.  SourcererCC performs much better then Deckard across all clone types.  While Deckard has decent recall for the C clones, its Java recall is very poor.  We believe this is due to its older Java parser (Java-1.4 only), while the Java reference clones may contain up to Java-1.6 features. 

In summary, SourcererCC has perfect recall with the Mutation Framework, which shows it can handle all the types of edits developers make on cloned code.  As per standard mutation analysis, the Mutation Framework only uses one mutation operator per clone.  This allows it to measure recall very precisely per type of edit and clone type.  It also prevents the code from diverging too far away from natural programming.  However, this means that the Mutation Framework makes simple clones.  It does not produce complex clones with multiple type of edits, and the Type-3 clones it produces generally have a higher degree of syntactical similarity.  To overcome this issue, we use the real-world benchmark BigCloneBench as follows.

\subsubsection{Recall Measured by BigCloneBench}
Here we measure the recall of SourcererCC using BigCloneBench and compare it to the competing tools.  We evaluate how its capabilities shown by the Mutation Framework translate to recall for real clones produced by real developers in real software-systems, spanning the entire range of clone types and syntactical similarity.  Together the benchmarks provide a complete view of SourcererCC's recall.

BigCloneBench~\cite{bcb_icsme14} is a Big Data clone benchmark of manually validated clone pairs in the inter-project software repository IJaDataset 2.0~\cite{ijadataset}.  IJaDataset consists of 25,000 open-source Java systems spanning 3 million files and 250MLOC.  This benchmark was built by mining IJaDataset for functions implementing particular functionalities.  Each clone pair is semantically similar (by their target functionality) and is one of the four primary clone types (by their syntactical similarity).  The published version of the benchmark considers 10 target functionalities~\cite{bcb_icsme14}.  We use an in-progress snapshot of the benchmark with 48 target functionalities, and 8 million validated clone pairs, for this study.

For this experiment, we consider all clones in BigCloneBench that are 6 lines or 50 tokens in length or greater.  This is a standard minimum clone size for benchmarking~\cite{bellon,bcb_icsme15}.  The number of clones in BigCloneBench, given this size constraint, is summarized per clone type in Table~\ref{tab:bcb_contents}.  There is no agreement on when a clone is no longer syntactically similar, so it is difficult to separate the Type-3 and Type-4 clones in BigCloneBench.  Instead we divide the Type-3 and Type-4 clones into four categories based on their syntactical similarity, as follows.  Very Strongly Type-3 (VST3) clones have a syntactical similarity between 90\% (inclusive) and 100\% (exclusive), Strongly Type-3 (ST3) in 70-90\%, Moderately Type-3 (MT3) in 50-70\% and Weakly Type-3/Type-4 (WT3/4) in 0-50\%.  Syntactical similarity is measured by line and by token after Type-1 and Type-2 normalizations.  We use the smaller of the measurements for categorization.  The categories, and the benchmark in general, are explained in more detail elsewhere~\cite{bcb_icsme14}.

\begin{table}
	\scriptsize 
	\centering
	\tabcolsep=0.10cm
	\caption{BigCloneBench Clone Summary}
	\label{tab:bcb_contents}
	\begin{tabular}{ccccccc}
		\toprule
		Clone Type & T1 & T2 & VST3 & ST3 & MT3 & WT3/T4 \\
		\midrule
		\# of Clone Pairs & 35787	& 4573	& 4156	& 14997	& 79756	& 7729291\\
		%				\cmidrule{2-7}
		%				& \multicolumn{6}{c}{7868560} \\
		\bottomrule
	\end{tabular}
	\vspace{-0.225in}
\end{table}

\textbf{Procedure.}  We executed the tools for IJaDataset and evaluated their recall with BigCloneBench.  As we saw previously (Section~\ref{sec:scale_eval}), most tools do not scale to the order of magnitude of IJaDataset (250MLOC).  Our goal here is to measure recall not scalability.  We avoid the scalability issue by executing the tools for a reduction of IJaDataset with only those files containing the known true and false clones in BigCloneBench (50,532 files, 10MLOC).  Some of the competing tools have difficulty even with the reduction, in which case we partition it into small sets, and execute the tool for each pair of partitions.  In either case, the tool is exposed to every reference clone in BigCloneBench, and it is also exposed to a number of false positives as well, creating a realistic input.  We measure recall using a subsume-based clone-matching algorithm with a 70\% threshold.  A tool successfully detects a reference clone if it reports a candidate clone that subsumes 70\% of the reference clone by line.  This is the same algorithm we use with the Mutation Framework, and is a standard in benchmarking~\cite{bellon}.

\textbf{Results.}  Recall measured by BigCloneBench is summarized in Table~\ref{tab:bcb_recall}.  It is is summarized per clone type and per Type-3/4 category for all clones, as well as specifically for the intra and inter-project clones.

\begin{table*}[t]
	\centering
	\scriptsize
	\tabcolsep=0.075cm
	\centering
	\caption{BigCloneBench Recall Measurements}\label{tab:bcb_recall}
	\begin{tabular}{ccccccccccccccccccccc}
		\toprule
		\multirow{2}{*}{Tool} & \multicolumn{6}{c}{All Clones} & & \multicolumn{6}{c}{Intra-Project Clones} & & \multicolumn{6}{c}{Inter-Project Clones} \\
		\cmidrule{2-7}\cmidrule{9-15}\cmidrule{16-21}
		& T1  & T2 & VST3 & ST3 & MT3 & WT3/T4 & & T1  & T2 & VST3 & ST3 & MT3 & WT3/T4 & & T1  & T2 & VST3 & ST3 & MT3 & WT3/T4 \\
		\midrule
		SorcererCC & 100 & 98  & 93  & 61 & 5  & 0 & & 100 & 99  & 99  & 86 & 14 & 0  & & 100 & 97  & 86  & 48 & 5 & 0 \\
		\midrule
		CCFinderX  & 100 & 93  & 62  & 15 & 1  & 0 & & 100 & 89  & 70  & 10 & 4  & 1  & & 98  & 94  & 53  & 1  & 1  & 0 \\
		Deckard    & 60  & 58  & 62  & 31 & 12 & 1 & & 59  & 60  & 76  & 31 & 12 & 1  & & 64  & 58  & 46  & 30 & 12 & 1 \\
		iClones    & 100 & 82  & 82  & 24 & 0  & 0 & & 100 & 57  & 84  & 33 & 2  & 0  & & 100 & 86  & 78  & 20 & 0 &  0 \\
		NiCad      & 100 & 100 & 100 & 95 & 1  & 0 & & 100 & 100 & 100 & 99 & 6  & 0  & & 100 & 100 & 100 & 93 & 1  & 0 \\
		
		\bottomrule
	\end{tabular}
	\vspace{-0.2in}
\end{table*}

SourcererCC has perfect detection of the Type-1 clones in BigCloneBench.  It also has near-perfect Type-2 detection, with negligible difference between intra and inter-project.  This shows that the 70\% threshold is sufficient to detect the Type-2 clones in practice.  SourcererCC has excellent Type-3 recall for the VST3 category, both in the general case (93\%) and for intra-project clones (99\%).  The VST3 recall is still good for the inter-project clones (86\%), but it is a little weaker.  SourcererCC's Type-3 recall begins to drop off for the ST3 recall (61\%).  Its recall is good in this Type-3 category for the intra-project clones (86\%) but poor for the inter-project clones (48\%).  We believe this is due to Type-3 clones from different systems having higher incidence of Type-2 differences, so the inter-project clones in the ST3 category are not exceeding SourcererCC's 70\% threshold.  Remember that the reference clone categorization is done using syntactical similarity measured after Type-2 normalizations, whereas SourcererCC does not normalize the identifier token names (to maintain precision and index efficiency).  Lowering SourcererCC's threshold would allow these to be detected, but could harm precision.  SourcererCC has poor recall for the MT3 and WT3/T4, which is expected as these clones fall outside the range of syntactical clone detectors~\cite{bcb_icsme15}.  Of course, Type-4 detection is outside the scope of this study.

Compared to the competing tools, SourcererCC has the second best recall overall, with NiCad taking the lead.  Both tools have perfect Type-1 recall, and they have similar Type-2 recall, with NiCad taking a small lead.  SourcererCC has competitive VST3 recall, but loses out in the inter-project case to NiCad.  SourcererCC is competitive with NiCad for intra-project clones in the ST3 category, but falls significantly behind for the inter-project case and overall.  NiCad owes its exceptional Type-3 recall to its powerful source normalization capabilities.  However, as we saw previously in Section~\ref{sec:scale_eval}, NiCad has much poorer execution time for larger inputs, and hits scalability constrains at the 100MLOC input.  So SourcererCC instead competes with execution performance and scalability, making these tools complimentary tools for different use-cases.

Comparison to CCFinderX is interesting as it is the only other tool to scale to the 100MLOC input.  Both tools have comparable Type-1 and Type-2 recall, with SourcererCC having the advantage of also detecting Type-3 clones, the most difficult type.  While BigCloneBench is measuring a non-negligible VST3 recall for CCFinderX, it is not truly detecting the Type-3 clones.  As shown by the Mutation Framework in Table~\ref{tab:mutationframework}, CCFinderX has no recall for clones with Type-3 edits, while SourcererCC has perfect recall.  Rather, CCFinderX is detecting significant Type-1/2 regions in these (very-strongly similar) Type-3 clones that satisfy the 70\% coverage threshold.  This is a known limitation in real-world benchmarking~\cite{jefficsme,bcb_icsme15}, which is why both real-world and synthetic benchmarking is needed.  CCFinderX's detection of these regions in the VST3 is not as useful to users as they need to manually recognize the missing Type-3 features.  CCFinderX's Type-3 recall drops off past the VST3 category, where Type-3 gaps are more frequent in the clones.  While we showed previously that CCFinderX also scales to larger inputs (Section~\ref{sec:scale_eval}), SourcererCC's faster execution, Type-3 support and better recall make it an ideal choice for large-scale clone detection.

Deckard and iClones are the other competing Type-3 clone detectors.  Both SourcererCC and iClones have perfect Type-1 recall, but SourcererCC exceeds iClones in both Type-2 and Type-3 detection, and iClones does not scale well.  Deckard has poor overall recall for all clone types, along with its scalability issues.

\subsection{Precision}
We measure SourcererCC's precision by manually validating a sample of its output.  We validate clones it detected for the BigCloneBench recall experiment so that the recall and precision measurements can be directly compared.  For the BigCloneBench input, SourcererCC detected 2 million clone pairs.  We randomly selected 390 of these clone pairs for manual inspection.  This is a statistically significant sample with a 95\% confidence level and a $\pm5$\% confidence interval.  We split the validation efforts across three clone experts.  This prevents any one judge's personal subjectivity from influencing the entire measurement.  The judges found 355 to be true positives, and 35 to be false positives, for a precision of 91\%.  This is a very strong precision as per the literature~\cite{6747168, taxonomy, roy:queens:07}, and demonstrates the accuracy and trustworthiness of SourcererCC's output.

The three judges made a common observation regarding the false positives.  These were cases where code fragments were syntactically similar, but not clones.  For example, unrelated but similar usage of a common API.  The false positives were not caused by a fault in SourcererCC, rather differentiating these instances from true clones is outside the scope of syntax-based clone detection.  We do not believe other syntax-based clone detectors would perform significantly differently.  In particular, a number of false positives were syntactically very similar unit test methods which were testing different behaviors of a class, but were both essentially a sequence of similar assertions.

Since the precision of the competing tools is publicly available (either by the tool authors and/or other experiments), and we wanted to give the maximum credit to them, we used those precision values for the comparison purpose.  For Deckard, a highest precision of 93\% was measured~\cite{deckard} using a higher similarity threshold.  However, Deckard's recall might be significantly lower in the setting of our study where we target Type-3 clones with an 85\% threshold.  The precision of CCFinder has been measured in two studies, and found to be 60-72\%~\cite{burdbailey,bellon}.  In general, token-based tools have lower precision~\cite{SMR:SMR520, bellon}.  This is caused by Type-2 normalizations, which causes false positives to have identical normalized token sequences.  Like CCFinder, iClones has poor precision due to the normalizations~\cite{bellon,SMR:SMR520}.  A study using artificial clones found a precision of 89-96\% for NiCad, depending on the configurations~\cite{mf_icstw09,mf_iwsc13}.  However, NiCad also exploits Type 2 normalizations including other advanced source transformations and the precision could drop significantly (e.g., around 60\%), depending on the configuration.

Compared to the competing tools, our tool has very strong precision at 91\%.  Unlike the competing token-based tools, we do not consider sequences of tokens, but rather the overlap of tokens on code blocks.  We do not normalize identifier names, so we are able to maintain a high precision.  Our recall results show that our overlap metric and threshold are able to provide high Type-2 and Type-3 recall without the identifier normalizations. 

\section{Threats of Validity}
\label{sec:threats}
As observed by Wang et al.~\cite{wang}, clone detection studies are affected by the configurations of the tools, and SourcererCC is no exception.  However, we carefully experimented with its configurations to achieve an optimal result.  As for the other tools, we also conducted test experiments, and also discussed with the corresponding developers for obtaining proper configurations, where available.  Their configurations also provided good results in our past studies~\cite{bcb_icsme15, jefficsme, jeff_scalability2}.  The measurement of precision is strongly affected by the personal subjectivity of the judge.  To combat this, we split the validation efforts across three clone experts so that the measurement reflects the views of multiple clone researchers.  For the same reason, for the competing tools, we compared with their published precision values (either by their tool authors and/or comparison experiments) instead of measuring them.

\section{Related Work}
\label{sec:rel-work}

Code clone detection is a mature area of study, and a number of detection techniques and tools have been presented in the literature.  Rattan et al.~\cite{rattan} found at least 70 clone detectors in the literature, almost double the 40 found by Roy et al.~\cite{taxonomy} in 2009.  However, very few tools target scalability to very large repositories.  In Section~\ref{sec:evaluation}, we compared SourcererCC with four state-of-the-art tools that were found to be competitive in previous studies~\cite{jefficsme,bcb_icsme15}, and thus we do not discuss them further here.

Liveri et al.~\cite{livieri:2007icse} introduced a method of distributing an existing non-scalable tool to very large inputs.  They partition the input into subsets small enough to be executed on a single machine, and execute the tool for each pair of partitions.  Partitioning achieved scalability in execution resource requirements, while scalability in time is achieved by distribution of the executions over a large number of machines.  Svajlenko et al.~\cite{jeff_scalability2} uses a similar methodology to scale the existing clone detectors.  Instead, the smaller inputs are chosen using a non-deterministic shuffling heuristic, which reduces the number of required tool executions significantly at the cost of some reduction in recall.  Distribution of these executions over a small number of machines is still recommended for scalability in time.  In contrast, SourcererCC uses a novel scalable clone detection technique, and is capable of scaling to large repositories on a single machine.

Ishihara et al.~\cite{ishihara} use hashing to scale method-clone detection for the purpose of identifying candidates for building libraries.  They use an AST-based parser to extract the methods and normalize their Type-1 and Type-2 features before producing a MD5 hash of their text.  Methods with the same hash-value are clustered into clone classes.  While they achieve fast execution time, their methodology does not detect Type-3 clones, which are the most common in large inter-project repositories~\cite{bcb_icsme14}.

Hummel et al.~\cite{hummel:icsm:2010} were the first to use an index-based approach to scale clone detection to large inter-project repositories.  Their technique detects only the Type-1 and Type-2 clones.  Their technique produces a very large index, so the index and the computation must be distributed using Google MapReduce.  In contrast, our SourcererCC produces a very small index, just 1.2GB for 18GB (250MLOC) of code, and detects Type-3 clones in large inter-project repository using a single machine.

Others have scaled clone detection in domain-specific ways, and are not directly related to ours.  Koshke~\cite{Koschke:CSMR:12} used suffix trees to scale license violation detection between a subject system and a large inter-project repository.  Keivanloo et al.~\cite{keivanloo:2011wcre} use an index approach to scale clone search to large inter-project repositories.  Chen et al.~\cite{Chen:2014:android} implement a technique for detecting cloned Android applications across large application markets.

\section{Conclusion}
\label{sec:conclusion}
In this paper, we introduced SourcererCC, a token-based accurate near-miss clone detection tool, that uses an optimized partial index and filtering heuristics to achieve large-scale clone detection on a standard workstation.  We demonstrated SourcererCC's scalability with IJaDataset, a Big Data inter-project repository containing 25,000 open-source Java systems, and 250MLOC.  We measure its recall using two state-of-the-art clone benchmarks, the Mutation Framework and BigCloneBench.  We find that SourcererCC is competitive with even the best of the state-of-the-art Type-3 clone detectors.  We manually inspected a statistically significant sample of SourcererCC's output, and found it to also have strong precision.  We believe that SourcererCC can be an excellent tool for the various modern use-cases that require reliable, complete, fast and scalable clone detection.  SourcererCC is available on our website~\footnote{\url{http://mondego.ics.uci.edu/projects/SourcererCC}}.

\section{Acknowledgement}
This material is partially supported by the National Science
Foundation grant No. 1218228, the DARPA MUSE Program and the
Natural Sciences and Engineering Research Council of Canada (NSERC).

%\bibliographystyle{abbrv}
%\newpage
\bibliographystyle{abbrv}

\bibliography{bib}  % sigproc.bib is the name of the Bibliography in this case
% ACM needs 'a single self-contained file'!
%\balancecolumns
%\balancecolumns % GM June 2007
% That's all folks!
\end{document}